\documentclass[twocolumn,prc,superscriptaddress,showpacs,amsmath,amssymb]{revtex4-1}

\usepackage{graphicx}
\usepackage{subfigure}
\usepackage{epsfig}
\usepackage{amsmath}
\usepackage{mathtools}
\usepackage{bm}

\usepackage{url}
\usepackage{multirow}
\usepackage{dcolumn}
\usepackage{amssymb}
\newcommand{\NN}{\mathbf{N}}
\newcommand{\MM}{\mathbf{M}}
\newcommand{\YY}{\mathbf{Y}}
\newcommand{\ulam}{\underline{\lambda}}
\newcommand{\uLam}{\underline{\Lambda}}
\newcommand{\uU}{\underline{U}}

\newcommand{\iso}[2]{$^{#2}$#1}

\begin{document}

% \begin{frontmatter}



\title{Method of Fission Product Beta Spectra Measurements for Predicting Reactor Anti-neutrino Emission}

\def\PNNL{Pacific Northwest National Laboratory, Richland, WA 99352, USA}

\author{D.M.~Asner}\affiliation{\PNNL}
\author{K.~Burns}\affiliation{\PNNL}
\author{L.W.~Campbell}\affiliation{\PNNL}
\author{B.~Greenfield}\affiliation{\PNNL}
\author{M.S.~Kos}\affiliation{\PNNL}
\author{J.L.~Orrell}\affiliation{\PNNL}
\author{M.~Schram}\affiliation{\PNNL}
\author{B.~VanDevender}\affiliation{\PNNL}
\author{L.S.~Wood}\affiliation{\PNNL}
\author{D.W.~Wootan}\affiliation{\PNNL}

\date{\today}

\begin{abstract}
The nuclear fission process that occurs in the core of nuclear reactors results in unstable, neutron-rich fission products that subsequently beta decay and emit electron antineutrinos. These reactor neutrinos have served neutrino physics research from the initial discovery of the neutrino to today's precision measurements of neutrino mixing angles. The prediction of the absolute flux and energy spectrum of the emitted reactor neutrinos hinges upon a series of seminal papers based on measurements performed in the 1970’s and 1980’s. The steadily improving reactor neutrino measurement techniques and recent reconsiderations of the agreement between the predicted and observed reactor neutrino flux motivates revisiting the underlying beta spectra measurements. A method is proposed to use an accelerator proton beam delivered to an engineered target to yield a neutron field tailored to reproduce the neutron energy spectrum present in the core of an operating nuclear reactor. Foils of the primary reactor fissionable isotopes placed in this tailored neutron flux will ultimately emit beta particles from the resultant fission products. Measurement of these beta particles in a time projection chamber with a perpendicular magnetic field provides a distinctive set of systematic considerations for comparison to the original seminal beta spectra measurements. Ancillary measurements such as gamma-ray emission and post-irradiation radiochemical analysis will further constrain the absolute normalization of beta emissions per fission. The requirements for unfolding the beta spectra measured with this method into a predicted reactor neutrino spectrum are explored.   
\end{abstract}



\maketitle

% \begin{keyword}


% \end{keyword}

% \end{frontmatter}

% \linenumbers


\section{Introduction}
\label{Introduction}

Neutrino experiments at nuclear reactors have played a vital role in the study of neutrino properties and flavor oscillation phenomenon.  The observed antineutrino rates at reactors are typically lower than model expectations~\cite{Huber2011,Mueller2011} . This observed deficit is called the “reactor neutrino anomaly”. Proposals exist for explaining this anomaly via non-standard neutrino physics models (sterile neutrinos, for example),  and a new understanding of neutrino physics may again be required to account for this deficit. However, model estimation uncertainties may also play a role in the apparent discrepancy. An experimental technique is proposed to make precision measurements of the beta energy spectrum from neutron induced fission using a 30 MeV proton linear accelerator~\footnote{For example the Project X Injector Experiment at Fermilab -see http://www-bdnew.fnal.gov/pxie/} as a neutron generator~\cite{Nagaitsev}. Each fission event produces fission products that decay and emit electrons (beta particles) and anti-neutrinos, and precise measurement of the beta energy spectrum is used to infer an associated anti-neutrino spectrum. The proposed new approach utilizes the flexibility of an accelerator-based neutron source with neutron spectral tailoring coupled with a careful design of an isotopic fission target and beta spectrometer. The inversion of the beta spectrum to the neutrino spectrum is intended to allow further reduction in the uncertainties associated with prediction of the reactor neutrino spectrum.

Through the fission process, four isotopes, $^{235}$U, $^{239}$Pu, $^{241}$Pu, and $^{238}$U contribute more than 99\% of all reactor neutrinos with energies above the inverse beta decay threshold (neutrino energy $\geq$ 1.8 MeV). The resulting predicted reactor neutrino flux is an accumulation of thousands of beta decay branches of the fission fragments. Reactor neutrino fluxes from the thermal fission of $^{235}$U, $^{239}$Pu, and $^{241}$Pu are currently obtained by inverting measured total beta spectra obtained in the 1980s at a beam port at the High Flux Reactor of the Institut Laue-Langevin (ILL) ~\cite{Huber2011}.  Recent reevaluations of the 1980s data with a careful investigation and treatment of the various sources of correlated and uncorrelated uncertainties indicated an upward shift of about 3\%, with uncertainties ranging from 2\% to 29\% across the neutrino spectrum ~\cite{Huber2011}. Clearly any limitations of the original ILL beta spectrum measurements in terms of energy resolution, absolute normalization, and statistical counting uncertainties will propagate into the predicted reactor antineutrino spectra. For a single beta decay branch, the neutrino energy spectrum is directly related to the beta energy spectrum by conservation of energy. However, there are hundreds of fission products and thousands of beta decay branches making measuring each branch individually practically impossible (especially for ultra-short half-life isotopes).  Thus measuring the cumulative beta spectrum remains the most viable technique for producing a representative spectrum used as a basis for inversion.  The beta spectrum from the fission target can be deconstructed into a set of individual beta decays modeled either as 'virtual branches'~\cite{Huber2011} or matched to expectations based on the information in nuclear decay databases. Likewise a parallel measurement of the gamma-ray emission from the irradiated fission foil (in situ and post-irradiation), provides a means to check the normalization of the beta emission per fission. These aspects of the proposed measurement seek to improve the confidence of the underlying reactor neutrino spectrum predictions.

\section{Neutron production and spectra}
\label{NeutronProduction}
\begin{figure}[ht]
\begin{center}
\includegraphics[height=0.36\textwidth]{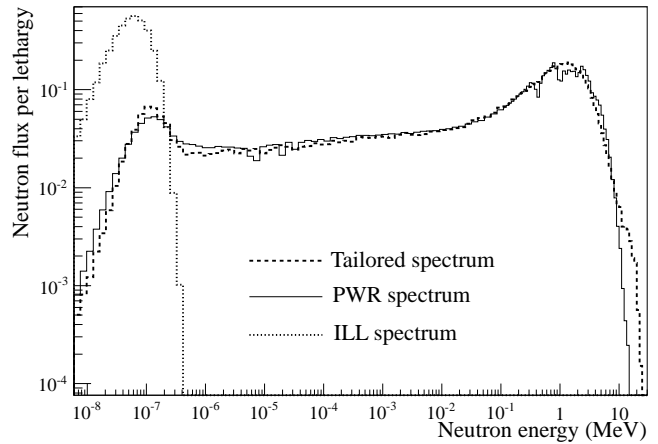}
\caption{\label{neutronspec} The neutron energy spectra from a PWR reactor,  D$_{2}$O thermalized neutrons (as at ILL),  and the tailored spectrum from the 30 MeV proton source.}
\end{center}
\end{figure}

The neutron spectrum in a nuclear reactor core is composed of three different energy ranges.  Neutrons from fission are emitted with an average energy of about 2 MeV and a most probable neutron energy of 0.73 MeV.  Figure~\ref{neutronspec} shows a representative neutron spectrum for a fuel pin in a pressurized water reactor (PWR).  In such a reactor, the fast portion of the neutron spectrum, with energies greater than 0.1 MeV,  has a shape similar to the primary fission neutrons.  In an operating reactor, fine structures in the neutron spectrum are introduced by absorption resonances on the fuel, moderator, and structural materials.  In the intermediate epi-thermal neutron energy range, from 0.1 MeV down to about 1 eV, the neutrons are slowing down with a characteristic 1/E dependence.  This is due to elastic scatters in the moderator removing a constant fraction of the neutron energy per collision (on average).  The thermal portion of the spectrum, below $\sim$1 eV, is characterized by a thermal Maxwellian flux shape, where the neutrons are in thermal equilibrium with the moderator.  The peak energy of the thermal flux depends upon the temperature of the moderator material.  Higher temperatures will shift the peak to higher energies.  At room temperature, the peak thermal flux is at 0.0265 eV, while for the PWR conditions in figure~\ref{neutronspec} the coolant is around 320 $^{\circ}\mathrm{C}$.  The magnitude and shape of the thermal spectrum depends on the relative volume fractions of moderator and fuel, and on the presence of burnable poisons or neutron absorbers inside the fuel or mixed in the moderator.  
The relative magnitudes of these three regions of the neutron spectrum depend a great deal on specific reactor conditions.  Neutron spectra at the beginning and end of an operating cycle will differ because of changing fuel isotopics from burnup, buildup of fission products, burnout of burnable poison in the fuel, and (in the case of PWRs) deliberate changes in the boron concentration in the coolant through the cycle.  The neutron spectra in various parts of the reactor core will vary because of increased leakage and/or reflection near the upper and lower surfaces and outer edges of the core compared to the interior of the core.  For boiling water reactors (BWRs), the coolant/moderator water density varies axially from full density water near the bottom of the core to full steam at the top of the core.

\begin{figure}[ht]
\begin{center}
\includegraphics[height=0.36\textwidth]{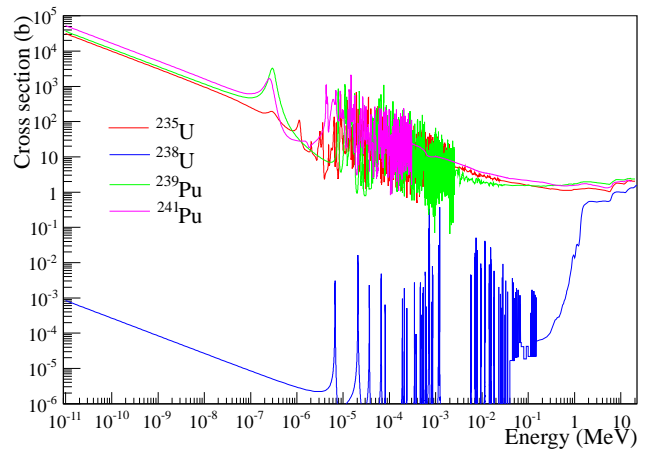}
\caption{\label{fissioncross} The fission cross-sections of key fuel isotopes.}
\end{center}
\end{figure}

\begin{figure}[htbp]
    \begin{center}
       \subfigure{\epsfig{file=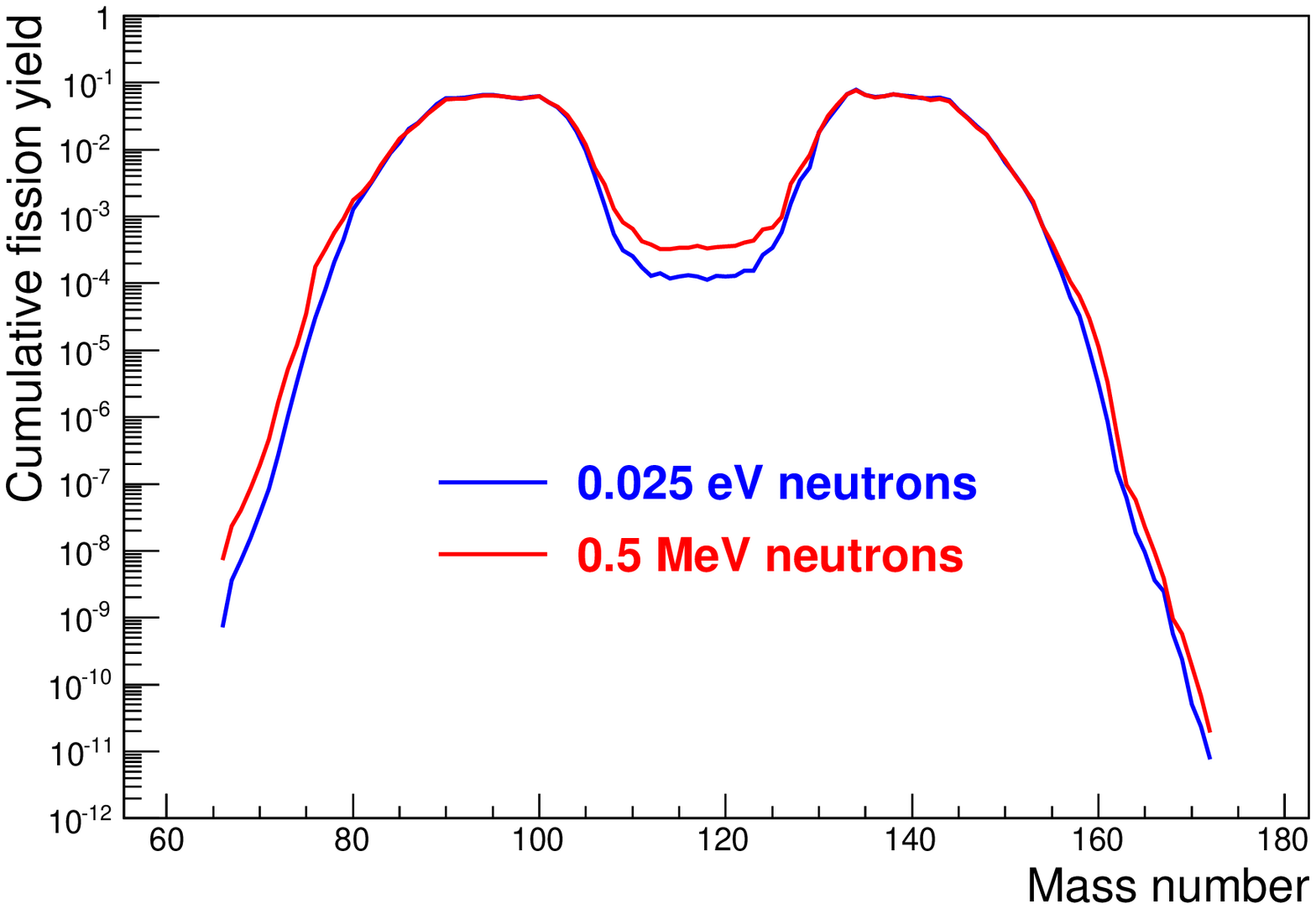, width=7.7cm}}
       \subfigure{\epsfig{file=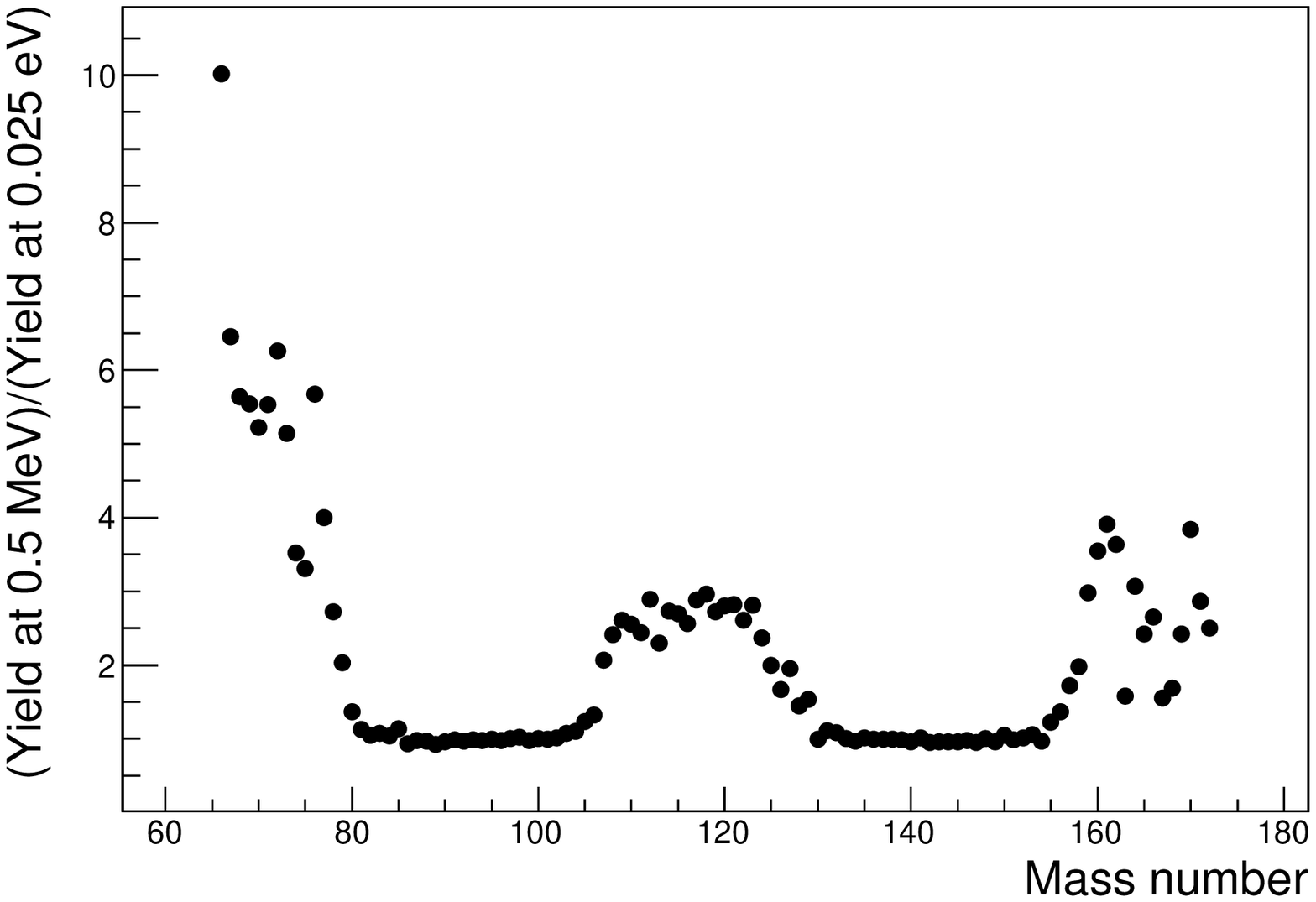, width=7.7cm}} 
       \caption[Neutron-induced fission yields by mass number for $^{235}$U fission.]{Cumulative neutron-induced fission yields by mass number for $^{235}$U fission for 0.025 eV incident neutron energies and for 0.5 MeV neutron energies (top panel).  Bottom panel shows the ratios of the cumulative yields for the two energies.  These plots illustrate the differences in fission yields for different incident neutron energies.} 
       \label{dproducts}
    \end{center}
\end{figure}

The primary fissioning isotopes in a typical commercial power reactor are $^{235}$U, $^{239}$Pu, $^{241}$Pu, and $^{238}$U.  Figure~\ref{fissioncross} shows the fission cross sections for these four isotopes.  The cross sections for $^{235}$U, $^{239}$Pu, and $^{241}$Pu are fairly flat at high energies, have a series of sharp resonances in the intermediate energy range, and a 1/v shape at thermal energies.  Both $^{239}$Pu and $^{241}$Pu have broad low energy resonances that reside at the transition between the thermal neutrons and the epi-thermal neutrons.  Uranium-238 has a threshold for fission at approximately 1 MeV, and therefore does not fission at lower energies.
Where the spectral variability impacts the neutrino anomaly is through the fission product yields.  There is a known dependence in fission product yields with the energy of the neutron causing fission.  This is illustrated in figure~\ref{dproducts}, which compares the fission product mass yields for thermal (0.025 eV) and 0.5 MeV neutrons for $^{235}$U fission.  Large differences can be seen for fission product masses in the central valley and for the lower and upper mass ranges.

A major advantage of an accelerator neutron source over a neutron beam from a thermal reactor is that the fast neutrons can be slowed down or tailored to approximate various power reactor spectra. This provides an advantage for control in studying how changes in the neutron spectra (i.e. in the reactor core) affects the resulting fission product beta spectrum. Furthermore, the $^{238}$U neutrino spectrum can be studied directly because of the enhanced 1 MeV fast neutron flux available at the accelerator source. Since $^{238}$U contributes on the order of 10\% of the fissions in a power reactor, measurement of the beta spectrum (and hence neutrino spectrum) should contribute to reducing the overall uncertainty in the reactor neutrino spectrum. 

Previous beta spectra measurements were conducted in the 1980’s by irradiating fission foils in a D$_{2}$O-moderated thermal flux tube external to the ILL reactor.  The neutron flux in that arrangement would be expected to be similar to that shown in figure~\ref{neutronspec}, which does not have any intermediate or fast neutron components. In order to reproduce a PWR  reactor spectrum, the temperature of the target moderator should be close to the temperature of the reactor moderator.  This can be a problem for water moderators, since high pressures are required to keep the water from boiling at PWR temperatures.  One way around this is to use metal hydride moderators that can maintain the hydrogen content at elevated temperatures, thereby matching the spectral shape of the in-core neutron flux.

The objective of spectral tailoring is to make the generated neutron spectrum look more like the reactor spectrum through the use of moderators and reflectors.  The major effort of the target studies was to see if an arrangement of proton beam target, moderator, and reflector could adequately simulate a representative PWR reactor neutron spectrum.  A series of parametric calculations with a very simple model were done to evaluate various target, moderator, and reflector materials and dimensions.  These results were then incorporated into more realistic models to further evaluate promising configurations.  The results are shown in figure~\ref{neutronspec}, which compares the spectrally tailored accelerator neutron spectrum with the PWR spectrum.  Good agreement can be seen between the two spectra at fast, intermediate, and thermal energy ranges.  When the $^{235}$U fission cross section is folded with the entire neutron spectrum, the effective one group cross section for the tailored accelerator spectrum was 38 barns, compared to 39 barns for the PWR spectrum.

\section{Proton beam and target design}
\label{ProtonBeam}
The beam of protons is produced by the linear accelerator and is characterized by proton energy, proton beam current, and beam profile at the target.  The beam current defines the number of protons per second striking the target.  The proton energy determines the reactions that produce neutrons and other particles in the target, and the depth of penetration.  The beam profile determines the areal energy deposition rate in the target.

The target converts the proton beam to neutrons through various (p,n) type reactions.  The number of neutrons produced per incident proton and the energy distribution of the neutrons depends on the material of the target.  The target also has to dissipate the energy deposited in it by the proton beam.  Therefore, material properties such as heat transfer coefficients are important.  A 30 MeV proton beam will deposit on the order of 20 kW of heat in the beam target.  A means for removing the heat deposited in the target must also be supplied. The thickness and shape of the target must be designed to accommodate the heat deposited within the target.  Since the 30 MeV protons penetrate only about 0.25 cm in the target, the heat and radiation damage is primarily in this thin surface layer exposed to the beam.  Altering the shape of the target to distribute the heating and damage over a larger surface is advisable.  This was investigated by using cone-shaped or wedge-shaped targets, and this appeared sufficient to limit target temperature to  $<$ 1000 $^{\circ}\mathrm{C}$ and maintain target integrity for several materials.

The neutrons generated in the target will generally have a distribution of energies up to the incident proton energy.  This spectrum of neutrons must be modified to mimic a reactor spectrum.  This can be done by including an adjustable length of moderator material for the neutrons to pass through.  In order to reproduce a reactor spectrum, the temperature of the target moderator should be close to the temperature of the reactor moderator.  A few centimeters of metal hydride moderator material at the temperature representative of PWR coolant conditions shows promise as the primary spectrum tailoring component.

\begin{figure}[ht!]
\begin{center}
\includegraphics[height=0.4\textwidth]{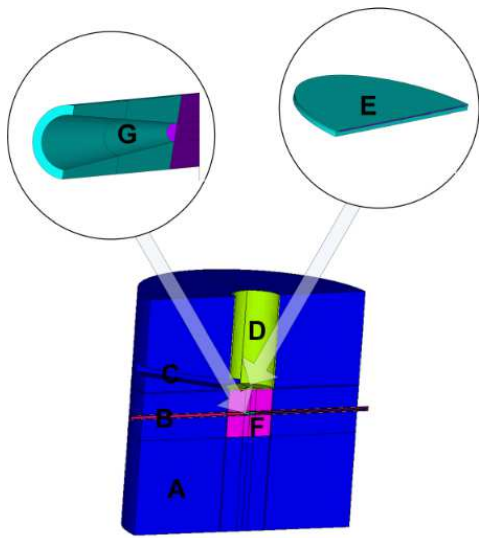}
\caption{\label{targetd} Schematic diagram of the proposed experimental setup.  The different labelled components are as follows:  A) lead reflector 40 cm diameter $\times$ 40 cm in height,  B) proton beam tube (0.5 cm in diameter), C) gamma window (see section~\ref{gammaconstraints}), D) beta tube (10cm in diameter), E) fission foil (see section~\ref{FissionFoil}), F) moderator, G) proton target.  The dimensions of some of the components may change in the final design.}
\end{center}
\end{figure}

A neutron reflector is proposed to reduce neutron leakage from the system, and to scatter neutrons released from the proton target back to the moderator and fission foil regions.  This reflector can also serve as a shield for the neutron, gamma, and other radiation generated in the proton target and fission foil.  Lead was found to be a good reflector.  Figure~\ref{targetd} shows a diagram of the proposed design.

\section{Fission foil}
\label{FissionFoil}
A fission foil will be placed in an area where the neutrons will have the desired spectrum.  Fission foils
of the primary fissioning isotopes will be used:  $^{235}$U, $^{239}$Pu, $^{241}$Pu, and $^{238}$U.  The neutron flux at the fission foil is predicted to be $>$ 10$^{11}$ n/cm$^{2}$/s,  with the tailored neutron spectrum.  It is estimated that the beta rates from the foil will be $>$ 10$^{8}$ betas/s per mg of $^{235}$U at an energy of $>$ 1 MeV.

Aspects of foil design such as density and composition that impact the emitted beta spectrum were investigated. Two primary design constraints guide the fission foil design for this experiment. The encapsulating foils must first serve to retain the highly radioactive fission products and secondly the foil must not significantly alter the outgoing beta spectrum to a degree that measurement quality is degraded. The retention of fission products and the preservation of the energy spectrum of emitted beta particles are opposing requirements.

Previous experiments like those conducted at ILL typically used nickel foils to encapsulate the fissionable material. Those experiments used nickel with an areal density of 7mg/cm$^{2}$, which translates to a thickness of approximately 7.85 $\mu$m.  The amount of fissionable material has varied between the past experiments.  The $^{241}$Pu foil used in~\cite{Hahn:1989vn} was 0.13 mg/cm$^{2}$ of 83\% enriched PuO$_{2}$ on a 2 $\times$ 6 cm$^{2}$ area.  For $^{235}$U,  0.15 mg/cm$^{2}$~\cite{Schreckenbach:1981fk} and 1 mg/cm$^{2}$~\cite{Schreckenbach:1985uq} of 93\% enriched UO$_{2}$ on a 3 $\times$ 6 cm$^{2}$ area have been used.

Primarily nickel foil and sputtered graphite have been investigated as candidates for materials to encapsulate the fissionable isotope(s). These two materials were chosen because they provide unique benefits that will allow prioritization of certain design criteria later in time. For example, the graphite can be made thinner, with an effective lower-Z and density, but costs more to fabricate.  

Preliminary Monte Carlo studies have been done of the beta spectra for various foil thicknesses. A thickness of 7.85 $\mu$m can be tolerated without significantly degrading the energy resolution of the entire system.  For the preliminary design,  similar foil dimensions and densities as those used in the ILL measurements will be used.

\section{Constraints on the beta spectra using gamma measurements}
\label{gammaconstraints}
Many beta decays are immediately followed by gamma radiation as the nucleus relaxes from an excited state to a ground or metastable state.  
This gamma radiation serves as a source of information with which to constrain the fission yields.  The feasibility of measuring the gammas produced by the decay of many of the fission daughters using a germanium detector situated some distance from the fission foil is considered.  The fission yields will then be obtained from the gamma measurements through a maximum likelihood fit in both energy and time.  

\subsection{Time dependent isotope populations}
\label{timedependentisotope}

Measuring the gammas produced during the decay of the fission daughter products can constrain fission yields.  However,  data that is in coincidence with the fission must be avoided since it will be dominated by prompt fission gammas and it will not be possible to resolve the necessary gamma energies required to determine fission yields.  This will be easier using an accelerator source compared to a continuous reactor source of neutrons since the fissions will occur only when the proton beam is on target.

The detector will be observing a population of isotopes within the uranium target which vary with time.  The population of an isotope will grow when it is produced via fission or fed by the decay of a parent isotope, and it will decrease when it decays.
If only a single isotope being produced by fission is considered, its population $N$ will be governed by the following rate equation
\begin{align}
\frac{dN}{dt} = f Y - \lambda N
\end{align}
where $f$ is the fission rate, $Y$ is the fission yield of the isotope, and $\lambda$ is the decay constant of the isotope.  

It is expected that that there will be periods when the proton beam will not be incident on target.  
The measurement time is divided into "windows" of piecewise constant reaction rate.
It may be chosen to have the detector not record events that occur during a given time window (for example, to ignore events while the beam is incident on the target in order to reduce the noise from neutron capture, neutron inelastic reactions, and so on). 
If time window $m$ starts at a time $t_m$, then at any time $t$ within the window there is the solution
\begin{align}
\label{rateeq1}
N(t) = \frac{f Y}{\lambda} \left( 1-e^{-\lambda(t-t_m)} \right) + N(t_m) e^{-\lambda (t-t_m)}.
\end{align}
It is known that the population of all fission products is zero at the beginning of the experiment, so that the population at the beginning of all windows can be built up by calculating $N(t_{m+1})$ once $N(t_m)$ is known.

It will also be important to know the time integral of the population
\begin{align}
D(t_1,t_2) = \int_{t_1}^{t_2} dt \, N(t).
\end{align}
If $t_1$ and $t_2$ are both in the same time window, then
\begin{align}
\label{drateeq1}
D(t_1,t_2) & = \frac{f Y}{\lambda^2} 
\left[
\lambda(t_2-t_1) - e^{-\lambda (t_1-t_m)} + e^{-\lambda (t_2-t_m)}
\right]
\\ \nonumber &\phantom{=}
+\frac{N(t_m)}{\lambda}
\left[
e^{-\lambda (t_1-t_m)} - e^{-\lambda (t_2-t_m)}
\right].
\end{align}
Also, the overall integrated population is defined as the sum of integrated populations over time windows in which the detector is recording.
\begin{align}
D = \sum_m^{\rm rec} D(t_m,t_{m+1})
\end{align}

In practice, there is a system of isotopes which decay into each other
\begin{align}
\label{popdiffeqij}
\frac{dN_i}{dt} = f Y_i - \sum_j \lambda_{ij} N_j
\end{align}
where $\lambda_{ii}$ is the decay constant for isotope $i$ and $-\lambda_{ij}/\lambda_{ii}$ is the branching ratio for isotope $j$ to decay to isotope $i$.
It is convenient to express collections of quantities related to each isotope, such as yields or populations, as column vectors in the space of isotopes which will be denoted with bold symbols.  
Matrices in this space will be denoted with an underline and the inner product by the dot-product symbol.  
In this notation, Eq. \ref{popdiffeqij} reads
\begin{align}
\label{popdiffeqvec}
\frac{d}{dt} \NN = f \YY - \ulam \cdot \NN.
\end{align}
This can be solved by diagonalizing the rate matrix
\begin{align}
\ulam = \uU \cdot \uLam \cdot \uU^{-1}
\end{align}
where $\uLam$ is the diagonal matrix of eigenvalues and $\uU$ is the matrix of right eigenvectors.
If a vector $\MM$ is defined such that 
\begin{align}
\label{Mvecdef}
\NN = \uU\cdot\MM,
\end{align}
then
\begin{align}
\frac{d}{dt} \MM = f \uU \cdot \YY - \uLam \cdot \NN.
\end{align}
These are a set of uncoupled equations for the scalars $M_i$, each of which can be solved using Eq. $\ref{rateeq1}$ and whose integrals can be found using Eq. $\ref{drateeq1}$.
This allows the solution of the coupled rate equations via Eq. \ref{Mvecdef}.  The vector $\YY$ is the quantity of interest.  It will be fit to the data by selecting a model that best represents the data.

\subsection{Fitting to data}

The data will be fit using a maximum likelihood minimization.  To speed up the minimization, a binned maximum likelihood fit is performed over discrete energy and time bins.  The following likelihood likelihood 
function is maximized
\begin{align}
\mathcal{F} = -n' \ln N + \sum_b^{\rm nbins} n'_b \ln W_b
\end{align}
where the weight of a bin $b$, $W_{b}$, is the sum of expected gamma and background contributions in that particular energy-time bin.  The parameter $n'$ is the total number of dead time corrected events, and $n'_b$ is the number of dead time corrected events in energy-time bin $b$.

Describing the methods of maximum likelihood estimation goes beyond the scope of this paper, as does the art of function minimization or maximization.
Interested parties are directed to the literature \cite{Green2011, Press1992}.
Note that while grouping the data may be less accurate, evaluation can also be much faster.
It may be worthwhile to pre-estimate the parameters using a binned analysis, and then finalize a solution using a maximum likelihood estimate on the individual events.

\subsection{Notable gamma-active isotopes}

The relative contribution of various isotopes to the gamma spectra will depend on the length of the irradiation and measurement periods, and to a lesser extent on the neutron spectrum and irradiated material.
Those lines with the strongest signals will be the most constrained by the data, allowing fits with lower uncertainty.
Taking and analyzing data in list mode will also provide constraints on the fission yields of parents of the gamma-active isotopes.
Table \ref{gammaactiveisotopes} lists many of the isotopes with strong gamma lines between 100 and 6000 keV which are expected to be observable.

\begin{table*}
\begin{tabular}{ l| l | l}
\iso{Pr}{148}  & \iso{Te}{136} \iso{I}{136m} \iso{I}{136} & \iso{Zr}{100} \iso{Nb}{100} \\
\iso{La}{146m} \iso{La}{146} \iso{Ce}{146} \iso{Pr}{146}  & \iso{Te}{135} \iso{I}{135} \iso{Xe}{135} & \iso{Y}{99} \iso{Zr}{99} \iso{Nb}{99m} \iso{Nb}{99} \\
\iso{La}{145} \iso{Ce}{145} \iso{Ba}{145} & \iso{Sb}{134m} \iso{Te}{134} \iso{I}{134} & \iso{Y}{98m} \iso{Y}{98} \iso{Nb}{98} \\
\iso{Ba}{144} \iso{La}{144} & \iso{Sb}{133} \iso{Te}{133m} \iso{Te}{133} \iso{I}{133}  & \iso{Y}{97m} \iso{Y}{97} \iso{Nb}{97m} \iso{Nb}{97} \\
\iso{Ba}{143} & \iso{Sb}{132m} \iso{Sb}{132}  & \iso{Sr}{96} \iso{Y}{96m} \\
\iso{Ba}{142} \iso{La}{142} & \iso{Sb}{131} \iso{Te}{131}  & \iso{Sr}{95} \iso{Y}{95} \\
\iso{Cs}{141} \iso{Ba}{141} & \iso{Tc}{106} & \iso{Rb}{94} \iso{Sr}{94} \iso{Y}{94} \\
\iso{Xe}{140} \iso{Cs}{140} & \iso{Tc}{104}  & \iso{Rb}{93} \iso{Sr}{93} \iso{Y}{93m} \\
\iso{Xe}{139} \iso{Cs}{139} \iso{Ba}{139} & \iso{Tc}{103}  & \iso{Kr}{92} \iso{Rb}{92} \iso{Sr}{92} \\
\iso{I}{138} \iso{Xe}{138} \iso{Cs}{138} & \iso{Nb}{102} & \iso{Kr}{91} \iso{Rb}{91} \iso{Sr}{91} \iso{Y}{91m} \\
\iso{I}{137} \iso{Xe}{137} & \iso{Zr}{101} \iso{Nb}{101} \iso{Mo}{101} \iso{Tc}{101} & \iso{Kr}{90} \iso{Rb}{90m} \iso{Rb}{90} \\
\end{tabular}
\caption{Prominent gamma-active isotopes from fission over time scales between several seconds and a day, with emissions between 100 and 6000 keV.  Isotopes are arranged by isobars, with daughter products to the right.}
\label{gammaactiveisotopes}
\end{table*}

\subsection{Detector placement}

In order to estimate fission yields from gamma spectra, it is necessary to collect sufficient data to provide a good fit with statistically meaningful results.
The effectiveness of a given instrument design and experimental irradiation and measurement schedule can 
be estimated by comparison to a previous analysis of beta-delayed gamma rays to determine fission yield.  
The analysis method described above was used in~\cite{PNNLREP} to extract fission yields from a set of data taken at Oregon State University's TRIGA reactor~\cite{Williford2013}.
This data set was taken from exposure of an \iso{U}{235} foil to a thermal beamline for 30 seconds and then measured for 150 seconds, repeated 100 times.  
As this data was primarily intended to demonstrate the analytical method, analysis was limited to the region between 3200 keV and 3650 keV.
The measurements recorded approximately $2\times 10^{6}$ gamma ray events in this energy range, of which approximately 350,000 were in the full energy peaks.  This provided sufficient data to determine relative fission yields of some of the most prominent isotopes in the spectrum (\iso{Cs}{142}, \iso{I}{137}, \iso{Y}{95}, \iso{Sr}{95}, \iso{Kr}{91}, \iso{Kr}{90}) within 10\% to 20\%.
It is expected that choosing measurement times more nearly equal to the irradiation times and performing the analysis over a wider spectral region would result in higher precision on the reported yields.

A GEANT4 \cite{Agostinelli2003, Allison2006} radiation transport simulation was set up to estimate a detector placement that would allow adequate statistics by comparison to the above experiment.  
An 8 cm diameter, 8 cm length HPGe detector was placed 100 cm from the fission foil.
A 20 cm thickness of borated polyethylene was placed around the reflector box to reduce neutron exposure to the detector.  
A narrow wedge-shaped viewing port was modeled into the neutron shield and reflector to allow the foil gamma rays to be observed by the detector.
An additional 10 cm slab of borated polyethylene was placed between the reflector and the detector, and 1 cm thickness of borated polyethylene was placed directly in front of the detector to reduce the dose from neutrons streaming down the viewing port.
In this geometry, the absolute peak efficiency in the 3200 to 3650 keV region was estimated to be $3.5\times 10^{-5}$.
To acquire comparable statistics $1\times 10^{10}$ gamma rays emitted from the uranium foil in the spectral region of interest during the collection times is required.

The model presented in section~\ref{timedependentisotope} allow analysis of the gamma rays per fission with a suitable data set for the fission yields and gamma intensities.  
Using ENDF \cite{ENDF} and ENDSF \cite{ENSDF} data sets, an estimate of a mean of $3.4\times 10^{-2}$ gamma rays in the 3200 to 3650 keV spectral region per fission event emitted during the measured time windows is obtained.
Given the expected mass of the fission foil, it is expected that the gamma flux will be low enough not to overwhelm the data acquisition while still being able to acquire the necessary statistics. 

This analysis suggests that a single high relative efficiency HPGe detector viewing gamma rays from the fission foil can achieve sufficient statistics for meaningful analysis without being overwhelmed by the rate of gamma interactions or destroyed by the neutrons.

\section{Beta spectrometer} 
\label{sec:betaSpectrometer}

A measurement of the fission foil beta spectrum needs to be made with good efficiency and energy resolution.  Efficiency needs to be good, and precisely known, since uncertainty in the energy dependent efficiency leads directly to a systematic uncertainty in predicted anti-neutrino flux.  Previous authors used magnetic beta spectrometry for the original fission beta specta measurements~\cite{Schreckenbach:1981fk, Feilitzsch:1982kx, Schreckenbach:1985uq, Hahn:1989vn, Haag:2013fk}.  Those measurements were made with a double focusing spectrometer named BILL~\cite{Mampe:1978zr}.  Electrons from the fission foil source were transported through a 14\,m long 10\,cm diameter beam tube to the spectrometer.  The spectrometer formed an image of the aperture slit on a pair of multi-wire proportional counters in the focal plane.  BILL was and still is an exquisite instrument, with relative momentum resolution of a few parts in $10^{4}$ for large targets and a momentum precision of one part in $10^5$.  

A preliminary design is explored where betas emitted from a fission foil activated as described in Section~\ref{ProtonBeam} are transported along a beam pipe to a simple dipole spectrometer with active tracking of betas performed by a time projection chamber (TPC)~\cite{Marx:1978,Hilke:2010} inserted between the pole faces.  There are several qualitative reasons for pursuing this design and analysis.  The active gaseous medium in the TPC is aninefficient detector of background gammas and neutrons.   Furthermore,  rare gamma and neutron interaction events can be rejected based on parameters of the TPC tracks.  It will be similarly easy to identify betas originating somewhere other than the source, betas from the source that have scattered in the beam pipe, and any other charged backgrounds.  Finally, this measurement method will provide a different set of systematic uncertainties to the measurements made with the BILL, ideally providing an "independent" test.  This section describes preliminary simulations of the performance.  Details of the TPC are given in Section~\ref{sec:TPC}.  Track reconstruction and overall performance are described in Section~\ref{sec:spectrometerPerformance}.  While the energy resolution of the TPC beta spectrometer will be worse than that of BILL,  in the next section it is demonstrated that the resolution of the TPC beta spectrometer is sufficient for extracting the anti-neutrino flux with an uncertainty of $<$ 1\%.

%\subsection{Electron transport to the spectrometer} \label{sec:betaTransport}

%Malachi?

\subsection{Tracking time projection chamber} \label{sec:TPC}

Most time projection chambers are large-volume devices, but smaller detectors do exist for specialized applications.  To simulate the response of a time projection chamber of appropriate size, the NIFFTE Fission TPC~\cite{Heffner:2005} was used as the baseline.  The NIFFTE TPC is designed to make precision cross section measurements of major actinides with an uncertainty of better than 1\%, and its design characteristics make it a good candidate for a beta spectrometer baseline design.

\begin{figure}
\begin{center}
\resizebox{0.35\textwidth}{!}{\includegraphics{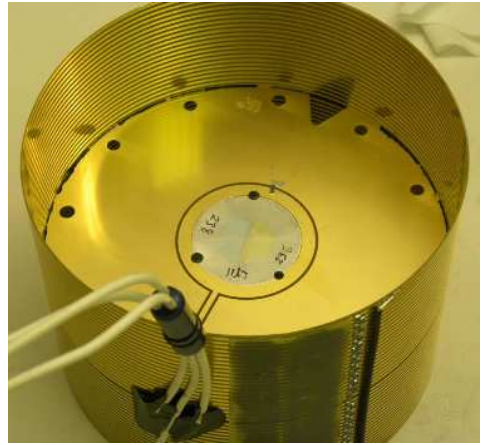}}
\caption{\label{fig:field_cage} Field cage and target cathode for the NIFFTE TPC.  The active volume is 5.4~cm deep and 15~cm across.}
\end{center}
\end{figure}

The NIFFTE TPC design consists of a cylinder 15~cm in diameter and 5.4~cm length (see figure~\ref{fig:field_cage}).  In the NIFFTE experiment two such volumes are used with a target placed between them to identify fission fragments exiting in both directions. For the beta spectrometer application only one side was simulated.  Each side is read out with a MICROMEGAS~\cite{Giomataris:1996} gain region and 2976 hexagonal readout pads of 2mm pitch.  The small drift volume allows for fast readout ($\approx$~1${\mu}s$ in P10 gas) and minimizes electron cloud diffusion.  The FPGA-based digital electronics are read out via Ethernet fiber to a central data acquisition computer.

The NIFFTE experiment design differs from the expected beta spectrometer design in several significant ways: only one TPC volume will be used; in NIFFTE the neutron beam enters the TPC axially, while in the beta spectrometer it will enter from the side; the NIFFTE experiment has no magnet, unlike the beta spectrometer application; and the signal gain in the NIFFTE MICROMEGAS is modest (10-40) since it is detecting fission fragments, while for the minimally ionizing betas in the spectrometer it will need to be significantly greater.  A triple GEM structure is a likely candidate for the actual gain stage~\cite{Sauli:1997} in the spectrometer application.

The simulation code for the NIFFTE TPC consists of a GEANT4~\cite{Agostinelli2003} particle transport and energy loss component, and a detector simulation component.  The detector simulation code handles effects such as electron cloud drift and diffusion, charge sharing between pads, preamplifier noise, and signal crosstalk.  For the beta spectrometer performance simulations, the output of the electron transport simulation was used as input to the TPC detector simulation.  Total charge collected on each pad was output for each event and passed on to the track identification code.

\subsection{Simulation of the Spectrometer} \label{sec:spectrometerPerformance}
The GEANT4 based simulation includes the entire path of the beta from the moment it exits the fission foil.  A significant degradation in energy resolution occurs due to scattering along the pipe from the fission foil to the spectrometer.  The final energy resolution reported at the end of this section includes any scattering that occurs during travel down the pipe.  The simulation concludes with the beta going through the spectrometer.  A track fitting algorithm is applied to determine its energy.
\subsubsection{Track identification and reconstruction}

\begin{figure}
\begin{center}
%\resizebox{0.8\textwidth}{!}{\includegraphics{track_5000G_10MeV.png}}
\resizebox{0.45\textwidth}{!}{\includegraphics{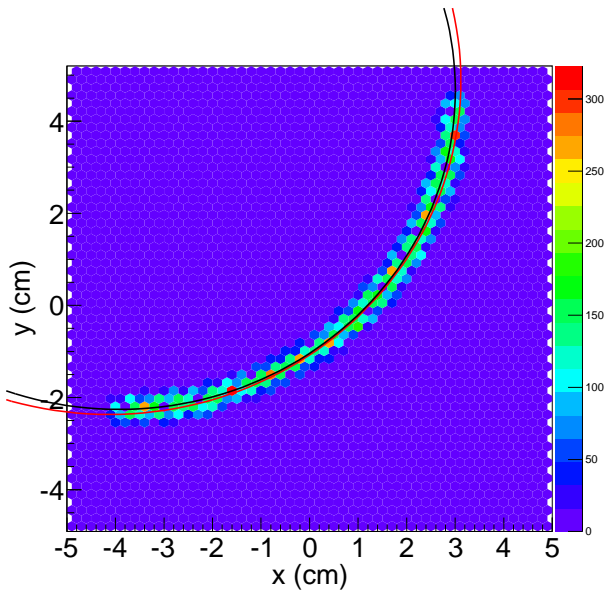}}
\caption{\label{fig:TPCTrack} A simulated TPC track for a 10\,MeV beta.  The color of each hexagonal pixel represent the collected charge above threshold in arbitrary uncalibrated units.  The red circular arc represents the initial track parameter guesses from a Hough transform.  The black circular arc represents the results of a continuous-parameter fit.}
\end{center}
\end{figure}

Track fitting and reconstruction is a two-step process.  The first step is to perform a circular Hough transformation on the recorded track.  That transformation is computationally intensive so the three-dimensional parameter space is divided into large bins representing values of the radius $r$ and the $(x,y)$-coordinates of the center $(a,b)$.  The results of this initial step are used as initial guesses for the parameters of a maximum likelihood fit allowing continuous values of the parameters.  A typical simulated track is shown in figure~\ref{fig:TPCTrack}.

The Hough transform uses recorded pixel charges to cast ``votes'' in a parameter space representing possible circular arcs of the form 
\begin{equation}\label{eqn:circle}
r^2 = (x-a)^2 + (y-b)^2.
\end{equation}
For each pixel $i$ centered at $(x_i, y_i)$ with a charge $q_i$ above threshold, a family of circles $r^2 = (a - x_i)^2 + (b - y_i)^2$ with all possible values of $a, b$ and $r$ is drawn in the Hough space.  A vote with weight $q_i$ is cast in every voxel intersected by each circle.  The voxel in the $(a, b, r)$ Hough space with the most votes corresponds to the most likely circular arc as in Eqn.~(\ref{eqn:circle}).

A quantity $\chi^2$ is computed as the value to be minimized in the continuous fit that refines the guess made by the Hough transform:
\begin{equation}\label{eqn:chi2}
\chi^2 = \frac{1}{Q^2} \sum_{i=0}^{N-1} \left( \frac{q_i d_i}{r_{\rm rms}} \right)^2,
\end{equation}
where the sum is over the $N$ pixels labeled $i = 0...N-1$ with charge above threshold, $q_i$ is the charge on the $i$th pixel, $d_i$ is the shortest perpendicular distance from the track to the center of the $i$th pixel, $r_{\rm rms}$ is the radius of the circle which contains 2/3 of the total area of a pixel, and $Q = \sum q_i$ is the total charge on all pixels above threshold.  The distance $d_i$ is just the length of the segment perpendicular to the track and passing through the pixel center $(x_i, y_i)$:
\begin{equation}\label{eqn:d_i}
d_i = \left| r - \sqrt{(x_i - a)^2 + (y_i - b)^2} \right|.
\end{equation}
Though the same notation is used, one should not interpret Eqn.~(\ref{eqn:chi2}) as the statistical parameter typically minimized in such a fit.  The relationship between the reconstructed track radius and the kinetic energy $E$ of a beta follows straightforwardly from relativistic kinematics:
\begin{equation}\label{eqn:E}
E = \sqrt{(eBrc)^2 + m^2 c^4} - mc^2,
\end{equation}
where $B$ is the magnitude of the uniform magnetic flux density in the tracking region, $m$ and $e$ are the electron mass and charge, respectively, and $c$ is the speed of light.

\subsubsection{Determining the overall performance of the spectrometer}

\begin{figure}
\begin{center}
%\resizebox{0.8\textwidth}{!}{\includegraphics{tc_10MeV.png}}
\resizebox{0.45\textwidth}{!}{\includegraphics{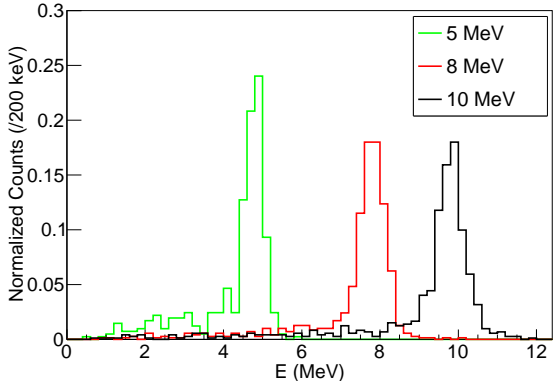}}
\caption{\label{fig:10MeVspectra} The spectra of 5, 8, and 10 MeV betas as measured by the tracking algorithm after transport through the beam pipe.}
\end{center}
\end{figure}

\begin{figure}
\begin{center}
%\resizebox{0.8\textwidth}{!}{\includegraphics{tcResVsE.png}}
\resizebox{0.45\textwidth}{!}{\includegraphics{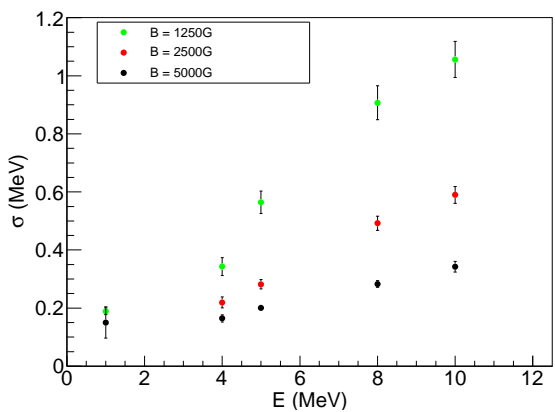}}
\caption{\label{fig:resVsE}  The fitted resolution, $\sigma$, versus true beta energy.  The error bars on the points are taken from the uncertainties on the fits.}
\end{center}
\end{figure}

Spectrometer performance is characterized by the simulation of monoenergetic beta particles emitted isotropically from the foil surface.  Scattering in the beam pipe is included but the effects of energy loss in the foil are not.  Figure~\ref{fig:10MeVspectra} shows the response to 5, 8, and 10 MeV  betas for a magnetic flux density of $B = 5000 G$.  
The spectra are all normalized to unit area.  Only events with $\chi^2 < 0.03$ are included.  Larger values of $\chi^2$ are indicative of multiply scattered events where the fitting algorithm fails since it is unable to distinguish more than one arc in an event.  These spectra are fit to a function of the form
\begin{equation}\label{eqn:lineShape}
f(E) = A \left( e^{\frac{(E-E_0)^2}{\sigma^2}} + \frac{m e^{\frac{E - E_0}{c}}}{1 + e^{\frac{E-E_0}{\sigma}}}  \right).
\end{equation}
The first term in Eqn.~(\ref{eqn:lineShape}) is Gaussian with mean $E_0$ and standard deviation $\sigma$.  The second term is a low side tail attributed to scattering in the beam pipe characterized by the additional parameters $m$ and $c$.  The spectra in figure~\ref{fig:10MeVspectra} and similar spectra for lower energy betas are fit with the form of Eqn.~(\ref{eqn:lineShape}).  Figure~\ref{fig:resVsE} shows the resolution $\sigma$ versus the energy for each of three magnetic flux densities.  Only fits with $p(\chi^2, {\rm NDF}) > 0.05$ are included.  

\section{Implications for an Anti-neutrino Analysis}
\begin{figure}[!htbp]
    \begin{center}
       \subfigure{\epsfig{file=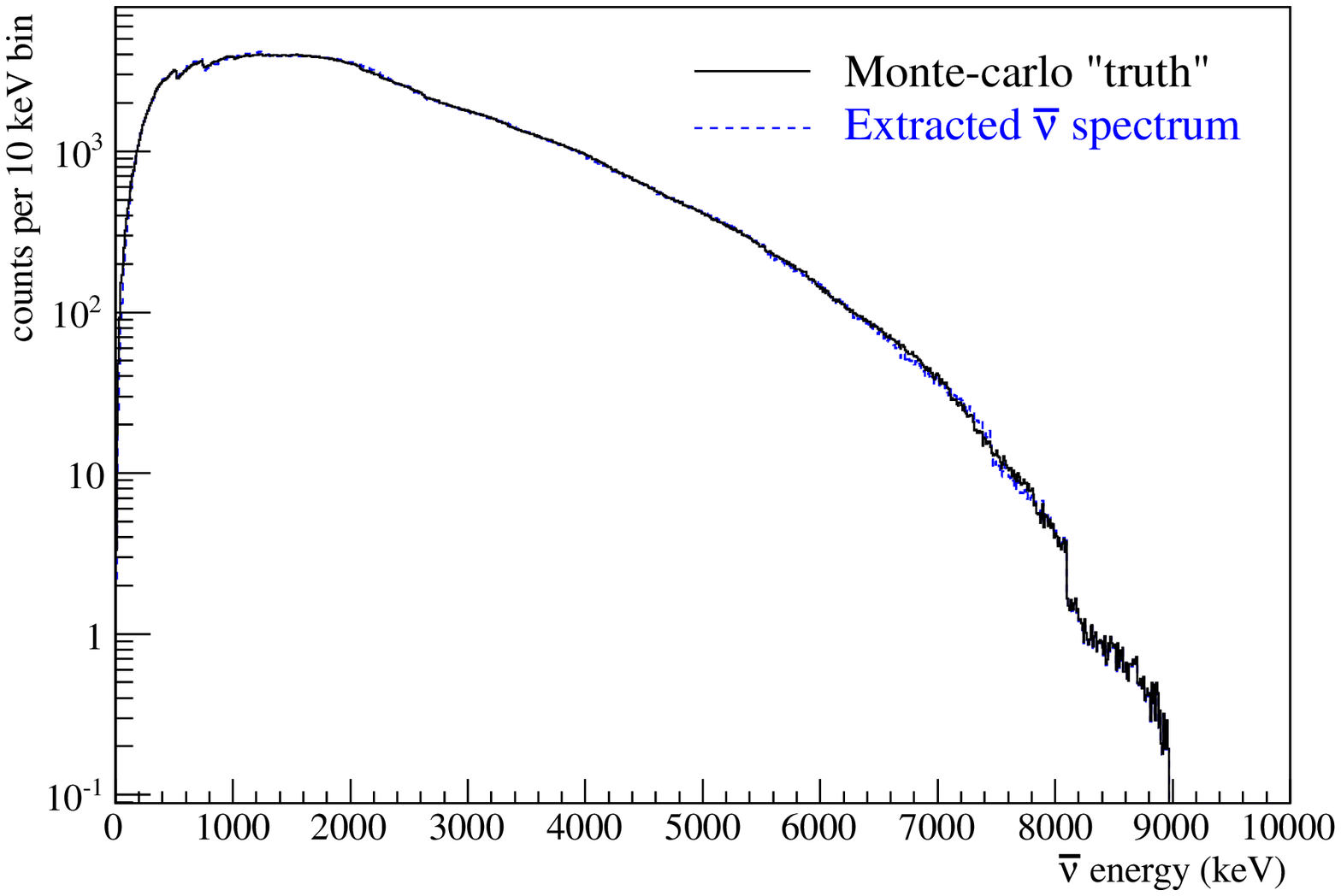, width=7.7cm}}
       \subfigure{\epsfig{file=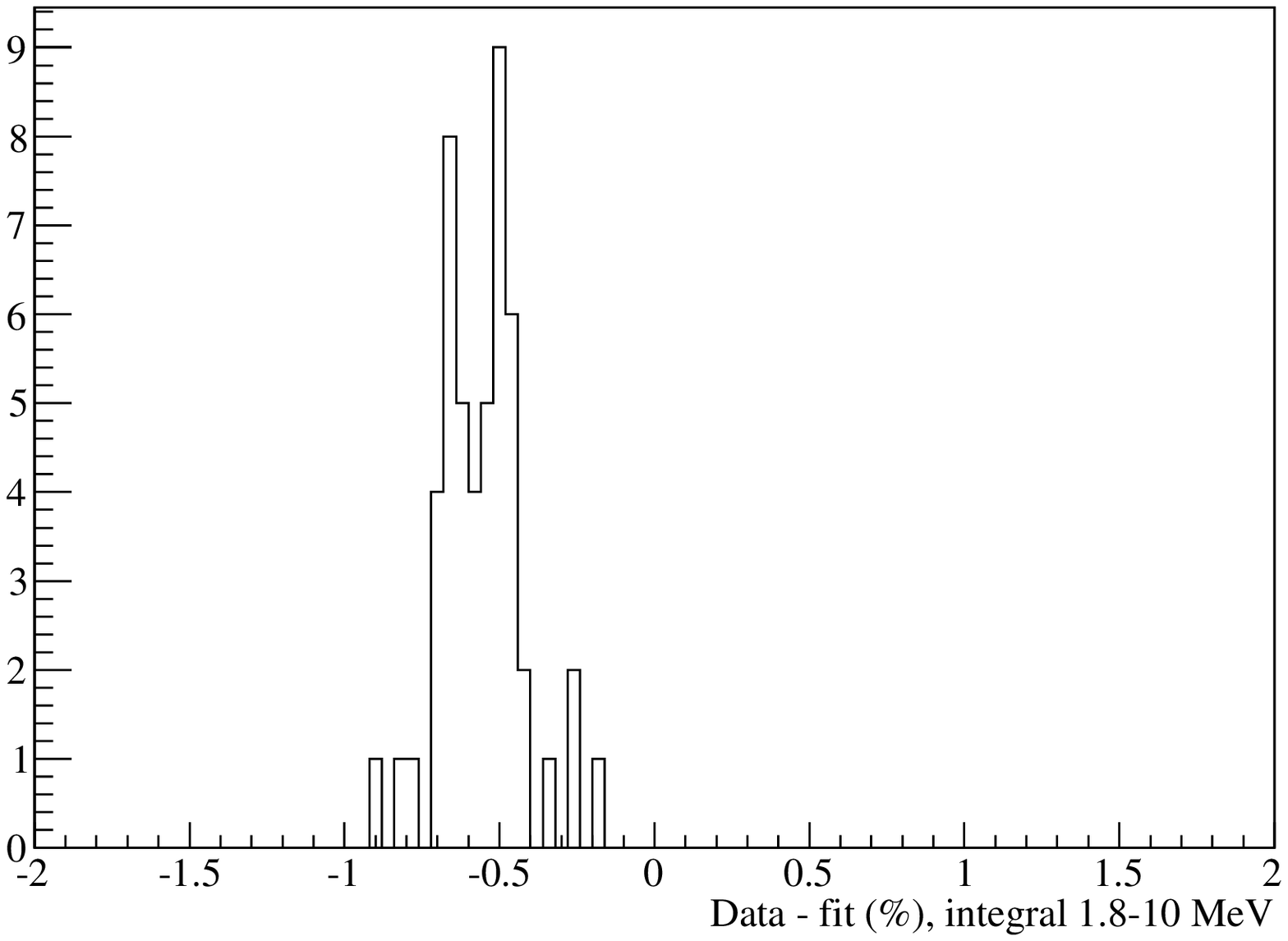, width=7.7cm}} 
       \caption[Histograms of the differences between the expected integrated anti-neutrino spectrum and the extracted anti-neutrino spectrum.]{Top panel shows a comparison of the extracted anti-neutrino spectrum and the ``true'' anti-neutrino spectrum given by the Monte Carlo for a 10\% energy resolution.  Bottom panel shows a histogram of the differences between the integrated true Monte Carlo and extracted anti-neutrino spectra for 50 signal extraction trials.} 
       
       \label{neutdiff}
    \end{center}
\end{figure}

The data from the beta spectrometer will ultimately be used to determine the corresponding anti-neutrino spectrum from the fission foil.  This will be done through a maximum likelihood signal extraction on the beta spectrum to determine the yields of the various fission product beta branches.  From the extracted yields the anti-neutrino spectrum is determined.  To quantify how the beta resolution affects the anti-neutrino spectrum a maximum likelihood signal extraction is performed on Monte Carlo of the $^{235}$U beta spectrum.  The Monte Carlo includes the corresponding anti-neutrino spectrum,  which is used to compare to the extracted anti-neutrino spectrum.  Various energy resolutions were tested in the Monte Carlo signal extraction.  As a worst case scenario,  a 10\% energy resolution is assumed, which is supported as a basis of estimation by figure~\ref{fig:10MeVspectra}.  With this worst case assumption figure~\ref{neutdiff} shows a comparison of the extracted anti-neutrino spectrum compared to the Monte Carlo's ``true'' anti-neutrino spectrum assuming a 10\% energy resolution.  With a 10\% energy resolution the uncertainty on the integrated anti-neutrino flux is expected to be $<$ 1\%.  As shown in the previous sections, the resolution of the beta measurements, including the scattering in the pipe, will be less than 10\%.    

The maximum likelihood signal extraction will also include measurements of the fission yields obtained from the proposed gamma analysis described in section~\ref{gammaconstraints}.  Also,  if possible,  radiochemical assays of the fission foil post irradiation will be used to obtain another set of measurements of the fission yields.  These independent measurements will be used as constraints in the signal extraction.  This will serve both to decrease the uncertainties on the extracted anti-neutrino spectrum as well as descrease the time required for the signal extraction to converge.

\section{Summary}
The persistence of the ``reactor neutrino anomaly''  warrants a new approach for measuring the 
beta spectra from fissionable material found in common nuclear reactors.  This paper outlines a plan
for using an accelerator neutron source coupled with a fission foil and a beta spectrometer to provide an 
independent measurement of the fission beta spectra.  The neutrons are produced through proton reactions 
on an appropriate target.  This approach is advantageous since the neutron spectrum can be tailored to be 
similar to the neutron spectra from different reactor types.  By careful study of target and moderator 
material a PWR neutron spectrum can be reproduced.  Simulations of a beta spectrometer, which relies on 
active tracking of betas in a TPC, show that the beta energy resolution of the system will allow 
measurements of the beta spectrum with the necessary precision to produce valuable constraints on the 
reactor anti-neutrino spectrum.  Furthermore,  independent measurements of the fission yields using 
germanium gamma spectroscopy and subsequent radiochemistry are planned.  These measurements will be used as external constraints in the maximum likelihood analysis to obtain the anti-neutrino spectrum.  
Details of an anti-neutrino spectrum extraction applied to the experimental setup described in the previous sections will be outlined in an upcoming paper.           		
\begin{acknowledgments}
The research described in this paper was conducted under the Laboratory Directed Research and Development Program at Pacific Northwest National Laboratory, a multiprogram national laboratory operated by Battelle for the U.S. Department of Energy under Contract DE-AC05-76RL01830.  
Detector simulations in this publication were based, in part, on the NIFFTE Fission Time Projection Chamber, and the authors would like to thank the NIFFTE Collaboration for use of their code.
\end{acknowledgments}

\bibliography{bib}






% \end{thebibliography}

\end{document}